\documentclass[aps,pra,preprint,endfloats, showpacs]{revtex4}
\pdfoutput=1 
\usepackage{amsmath}
\usepackage{amssymb}
\usepackage{graphicx}

\begin{document}

\title{Production of very-high-$n$ strontium Rydberg atoms }

\author{S. Ye}

\author{X. Zhang}

\author{T. C. Killian}

\author{F. B. Dunning}

\affiliation{Department of Physics and Astronomy and the Rice Quantum Institute,
Rice University, Houston, TX 77005-1892, USA}

\author{M. Hiller$^{1,2}$}

\author{S. Yoshida$^1$}

\author{S. Nagele$^1$}

\author{J. Burgd\"{o}rfer$^1$}

\affiliation{$^1$ Institute for Theoretical Physics, 
Vienna University of Technology, Vienna, Austria, EU}
\affiliation{$^2$ Physikalisches Institut, 
Albert-Ludwigs-Universit\"at Freiburg, Freiburg, Germany, EU}

\begin{abstract}
The production of very-high-$n$, $n\sim300$-500, strontium Rydberg
atoms is explored using a crossed laser-atom beam geometry. $n$$^{1}$S$_{0}$
and $n$$^{1}$D$_{2}$ states are created by two-photon excitation
via the 5s5p $^{1}$P$_{1}$ intermediate state using radiation with
wavelengths of $\sim$~461 and $\sim$~413~nm. Rydberg atom densities
as high as $\sim 3 \times 10^{5}$~cm$^{-3}$ have been achieved, sufficient
that Rydberg-Rydberg interactions can become important. The isotope
shifts in the Rydberg series limits are determined by tuning the 461~nm
light to preferentially excite the different strontium isotopes. 
Photoexcitation
in the presence of an applied electric field is examined. The initially
quadratic Stark shift of the $n$$^{1}$P$_{1}$ and $n$$^{1}$D$_{2}$
states becomes near-linear at higher fields and the possible use of
$n{}^{1}$D$_{2}$ states to create strongly-polarized, quasi-one-dimensional
electronic states in strontium is discussed. The data are analyzed
with the aid of a two-active-electron (TAE) approximation. The two-electron
Hamiltonian, within which the Sr$^{2+}$ core is represented by a semi-empirical
potential, is numerically diagonalized allowing calculation of the
energies of high-$n$ Rydberg states and their photoexcitation probabilities. 
\end{abstract}

\maketitle

\section{Introduction }

Studies of alkali atoms in states with large principal quantum numbers
$n$, $n\ge300$, have demonstrated the remarkable precision with
which atomic Rydberg states can be controlled and manipulated using
one, or more, electric field pulses, and have provided a wealth of
new insights into the dynamics of chaotic systems and into physics
in the ultra-fast ultra-intense regime~\cite{dunn09,buch02,jens91}.
These studies, however, have been limited to singly-excited electronic
states by the difficulty of exciting a second electron in the closed-shell
core ion. This restriction can be removed by use of alkaline-earth
atoms which have two valence electrons because, following excitation
of one electron to a high-$n$ state, the presence of the second valence
electron leaves a readily-excited optically-active core ion. For low
(total) angular momentum (low-$L$) Rydberg states, excitation of
the core ion leads to rapid autoionization through electron-electron
scattering~\cite{mill10,cook78}. As $L$ increases the autoionization
rate decreases and this decrease can serve as a probe of the evolution
of the Rydberg electron towards high-$L$ states. For sufficiently
high $L$ the core ion behaves as an independent, essentially-free
particle allowing it (and the Rydberg atom) to be manipulated through
optical trapping or imaged through laser-induced fluorescence~\cite{mcqu13}.
The autoionization process has been studied by colliding two radially
localized electronic wave packets in barium revealing violent rather
than gradual energy transfer between the two electrons~\cite{pish04}.
Furthermore, the presence of the second valence electron also admits
the possibility of creating two quasi-stable two-electron excited
states in the planetary atom~\cite{kali03} 
or frozen planet configurations~\cite{eich90,eich92,rich90}.
By taking advantage of the different energy-level structure of singlet
and triplet excited states, attractive \textit{and} repulsive inter-atomic
interactions in ensembles of Rydberg atoms can be realized using the
same element~\cite{vail12}. The goal of our ongoing experimental
and theoretical efforts is thus to explore the opportunities that
strontium provides to probe new aspects of Rydberg atom physics related
to two-electron excited states created either within a single atom
or by forming pairs of strongly-coupled Rydberg atoms with well-defined
interatomic separation. Central to these studies is the production
of very-high-$n$ states as these can be readily manipulated using
pulsed electric fields~\cite{dunn09} and have extraordinarily strong
long-range interactions. With these applications in mind we examine
here the creation of $n\sim$~300-500 strontium Rydberg atoms via
two-photon excitation from the ground state using a crossed atom-laser
beam geometry. In the first part of this study, we explore the creation
of strontium $n$$^{1}$S$_{0}$ and $n$$^{1}$D$_{2}$ Rydberg states.
Large photoexcitation rates are achieved that permit creation of multiple
Rydberg atoms with internuclear separations approaching those at which
Rydberg-Rydberg interactions become important. This opens up the opportunity
to study strongly-coupled Rydberg-Rydberg systems under carefully
controlled conditions. As a test of the spectroscopic resolution achievable
we obtain the isotope shifts in the series limits for the different
strontium isotopes present in the beam. Another focus is on photoexcitation
in a dc electric field. The character of the optically accessible
Rydberg states changes markedly as the strength of the applied field
is increased. We examine these changes and discuss the possibilities
for generation of strongly polarized states, which form the basis
of many of the protocols used to engineer Rydberg wave packets. The
results are analyzed with the help of a two-active-electron (TAE)
model that employs numerically exact diagonalization of the two-electron
Hamiltonian containing an empirical potential representing the Sr$^{2+}$
core. Exploiting the $n$ scaling for Rydberg states allows the comparison
of converged numerical results obtained near $n=50$ to data recorded
experimentally for $n\gtrsim300$.

\section{ Experimental approach }

\begin{figure}[p]
\includegraphics[width=15cm]{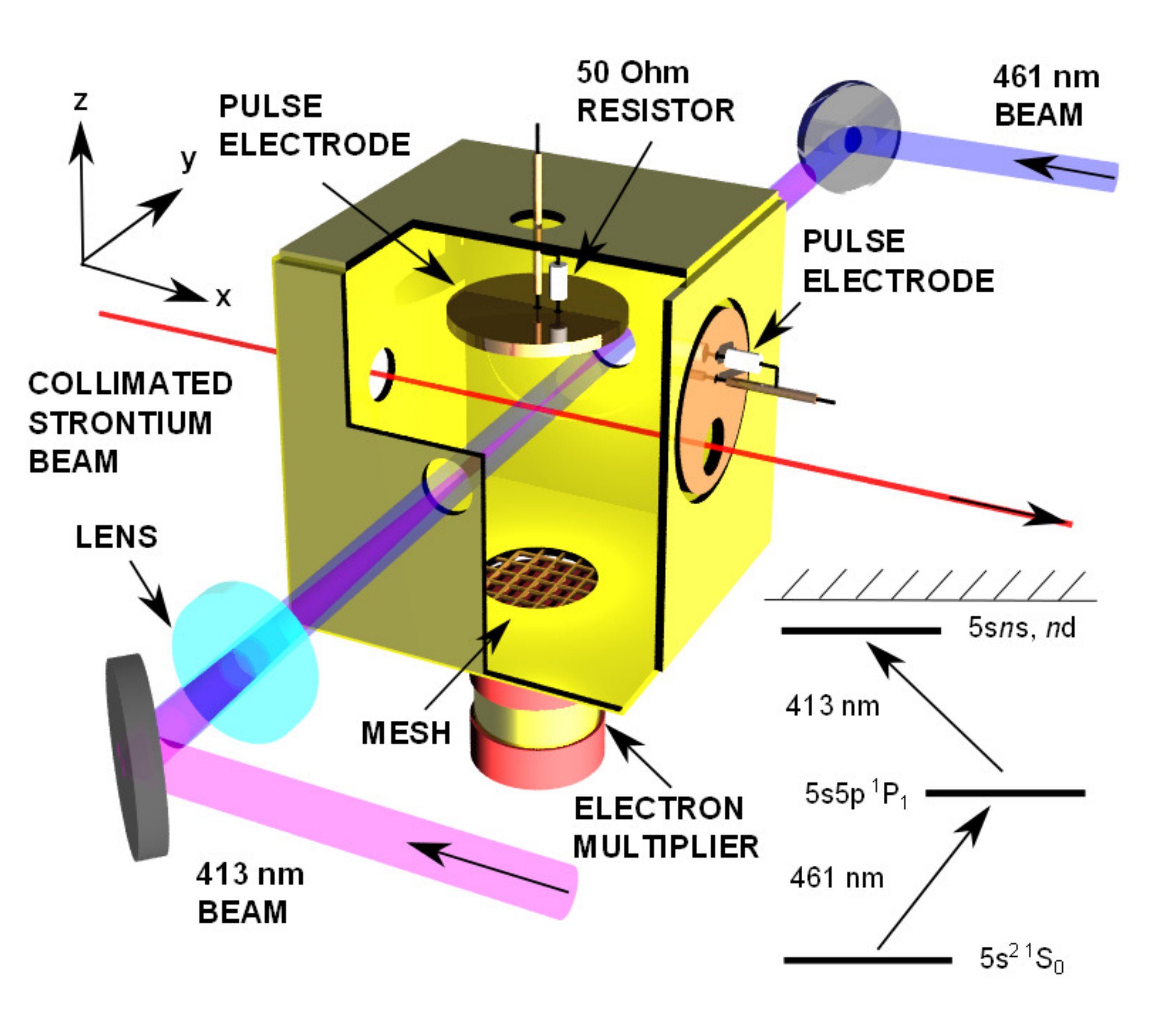}
\caption{\label{fig:apparatus} (Color online) Schematic diagram of the apparatus.
The inset shows the two-photon excitation scheme employed. }
\end{figure}

The experimental apparatus is shown in Fig.~\ref{fig:apparatus}.
Strontium atoms contained in a tightly-collimated beam are excited
to the desired high-$n$ (singlet) state using the crossed outputs
of two frequency-doubled diode laser systems. The first ``blue''
461~nm laser is tuned to the 5s$^{2}$ $^{1}$S$_{0}$ $\to$ 5s5p
$^{1}$P$_{1}$ transition while the second ``purple'' 413~nm laser
drives the transition from the intermediate 5s5p $^{1}$P$_{1}$ state
to the target state (this two-photon scheme is shown in the inset
of Fig.~\ref{fig:apparatus}). When crossing the atom beam, the linearly
polarized laser beams travel in opposite directions. By means of half-wave
plates, their polarization vectors can be aligned with either the
$x$- or $z$-axis. We explore excitation with both collinear ($z$-$z$)
as well as orthogonal ($x$-$z$) laser polarizations. Since the two
wavelengths are comparable, the use of counter-propagating light beams
can largely cancel Doppler effects associated with atom beam divergence
resulting in narrow effective experimental line widths. The strontium
atom beam is provided by an oven that can operate at temperatures
of up to 680$^{\circ}$C and which can, with appropriate collimation,
provide a $\sim$1mm-diameter beam with a full width at half maximum
divergence of $\sim$ 4~mrad at densities approaching 10$^{9}$~cm$^{-3}$.
As described elsewhere~\cite{frey93}, residual stray fields in the
experimental volume are reduced to $\le50\mu$Vcm$^{-1}$ by application
of small offset potentials to the electrodes that surround it.

Measurements are conducted in a pulsed mode. The output of the 461~nm
laser is chopped into a series of pulses of $\sim$~0.5~$\mu$s
duration and 20~kHz pulse repetition frequency using an acousto-optical
modulator. The laser is unfocused and has a diameter of $\sim$~3~mm.
Its intensity, $\sim$~10~mW~cm$^{-2}$, was selected to limit
line shifts and broadening due to effects such as the ac Stark shift
and Autler-Townes splitting~\cite{autl55}. Its pulse width, $\sim$~0.5~$\mu$s,
is chosen because for shorter pulse durations the widths of the spectral
features become increasingly transform limited. The 413~nm laser
remains on at all times and its beam is focused to a spot with a full
width at half maximum (FWHM) diameter of $\sim$~170~$\mu$m, resulting
in an intensity of $\sim$~250~W~cm$^{-2}$. The frequencies of
both lasers are stabilized and controlled with the aid of an optical
transfer cavity locked to a polarization-stabilized HeNe laser. This
cavity allows uninterrupted tuning of the lasers over frequency ranges
of up to $\sim$~800~MHz. Following each laser pulse, the probability
that a Rydberg atom is created is determined by state-selective field
ionization for which purpose a slowly-rising (risetime $\sim$~1~$\mu$s)
electric field is generated in the experimental volume by applying
a positive voltage ramp to the lower electrode. Product electrons
are accelerated out of the interaction region and are detected by
a particle multiplier. Because only one output pulse from the multiplier
can be detected following each laser pulse in the present setup, the
probability that a Rydberg atom is created during any laser pulse
is maintained below $\sim$~0.4 to limit saturation effects. This
can be accomplished by reducing the strontium atom beam density by
operating the oven at a lower temperature ($\sim$500$^{\circ}$C),
and/or by reducing the 413~nm laser power using a neutral density
filter.

\section{Two-active-electron model }

\subsection{ Model Hamiltonian }

While alkali atoms are well described by single-active electron models,
the same does not hold for alkaline-earth atoms which, in addition
to the closed-shell configuration of core electrons, possess two valence
electrons. We analyze the excitation spectra for strontium by means
of a two-active-electron (TAE) model. The Hamiltonian reads 
\begin{equation}
H=\frac{p_{1}^{2}}{2}+\frac{p_{2}^{2}}{2}+V_{\ell_{1}}(r_{1})
+V_{\ell_{2}}(r_{2})+\frac{1}{|\vec{r}_{1}-\vec{r}_{2}|}\,,\label{eq:hamil}
\end{equation}
 where $V_{\ell_{i}}(r_{i})$ is an angular-momentum dependent semi-empirical
model potential~\cite{ayma96} representing the Sr$^{2+}$ ion. The
latter includes a core polarization correction and is of the form
\begin{eqnarray}
V_{\ell}(r) & = & -\frac{1}{r}\left[2+36\exp(-\alpha_{1}^{\ell}r)
+\alpha_{2}^{\ell}\, r\,\exp(-\alpha_{3}^{\ell}r)\right]\nonumber \\
&&
 -\frac{\alpha_{{\rm cp}}}{2r^{4}}
\left[1-\exp[-(r/r_{c}^{\ell})^{6}]\right]\,,
\label{eq:modelpot}
\end{eqnarray}
where the parameters, $\alpha_{i}^{\ell},r_{c}^{\ell}$, are obtained
by fitting to known energy levels of the Sr$^{+}$ ion and the experimental
core polarizability $\alpha_{{\rm cp}}=7.5$ is $\ell$-independent~\cite{ayma96}.
(Atomic units are used throughout, unless otherwise stated.) Fine-structure
(FS) corrections are neglected as the preliminary calculations
including FS corrections show that they are of minor
importance for very high $n$. The atomic eigenenergies and eigenstates 
are obtained by numerical diagonalization of the Hamiltonian 
(Eq.~\ref{eq:hamil}). First, the single-particle
orbitals $|\phi_{n_{i},\ell_{i},m_{i}}\rangle$ and orbital energies
$E_{n_{i},\ell_{i},m_{i}}$ for the Sr$^{+}$ ion with 
\begin{equation}
H_{{\rm ion}}=\frac{p^{2}}{2}+V_{\ell}(r)\label{eq:hion}
\end{equation}
are generated using the generalized pseudo-spectral method~\cite{tong97}.
This method can efficiently describe wave functions on both long and short length scales by 
using a non-uniform spatial grid optimized for the potential employed. When FS corrections are 
included, the orbital energies agree well with the measured values~\cite{lang91,sans12}.
Whereas spin-orbit coupling is neglected (Eq.~\ref{eq:hion}), the calculated 
energies do agree with the spin-averaged measured values~\cite{ayma96}.
However, measured energy levels for the ionic p-states are available only for $n_i \le 8$ 
which introduces uncertainty in the the model potential parameters.
Using these one-electron orbitals the basis states of the two-electron
Hamiltonian (Eq.~\ref{eq:hamil}) are constructed as 
\begin{eqnarray}
|n_{1}\ell_{1}n_{2}\ell_{2};LM\rangle
=\sum_{m_{1}+m_{2}=M}
\frac{1}{\sqrt{2}}
\Big[
  C(\ell_{1},m_{1};\ell_{2},m_{2};L,M)| \phi_{n_{1},\ell_{1},m_{1}}\rangle
  |\phi_{n_{2},\ell_{2},m_{2}}\rangle\nonumber \\
  \pm (-1)^{\ell_1+\ell_2+L}
  C(\ell_{2},m_{2};\ell_{1},m_{1};L,M)| \phi_{n_{2},\ell_{2},m_{2}}\rangle
  |\phi_{n_{1},\ell_{1},m_{1}}\rangle
\Big]\,,
\label{eq:basis2e}
\end{eqnarray}
where $L$ is the total angular momentum, $M$ is its projection
onto the quantization axis, and the Clebsch-Gordan coefficients are
given in terms of 3j symbols as 
\begin{eqnarray}
C(\ell_{1},m_{1};\ell_{2},m_{2};L,M)=(-1)^{-\ell_{1}+\ell_{2}-M}\sqrt{2L+1}
\nonumber \\
\times\begin{pmatrix}\ell_{1} & \ell_{2} & L \\
m_{1} & m_{2} & -M
\end{pmatrix}\,.
\end{eqnarray}
The $\pm$ sign in Eq.~(\ref{eq:basis2e}) distinguishes basis states
symmetric (antisymmetric) with respect to electron exchange in coordinate
space representing the singlet (triplet) sector. In order to obtain
the spectrum in the presence of a dc electric field oriented along
the $z$-axis, the Hamiltonian $H(F)=H+F(\hat{z}_{1}+\hat{z}_{2})$
with the two-electron dipole operator $\hat{z}_{1}+\hat{z}_{2}$ is
diagonalized, yielding the desired eigenvalues and eigenstates.

Since the principal quantum numbers $n_{1},n_{2}$ of the orbitals
describe the excited states of the Sr$^{+}$ ion and not those of
neutral strontium, the quantum number $n$ of the Rydberg atom associated
with the calculated eigenstates of the two interacting electrons 
\begin{equation}
|nLM\rangle= \sum_{n_{1},\ell_{1}} \sum_{n_{2},\ell_{2}}
c_{n_{1},n_{2},\ell_{1},\ell_{2}} |n_{1}\ell_{1}n_{2}\ell_{2};LM\rangle \,.
\label{eq:basisRyd}
\end{equation}
is assigned in agreement with known excitation series taking into
account perturber states. This is not always straightforward as a
few states are hard to identify. For example, the spectroscopic literature
\cite{sans12} lists a 4d5p state in the $^{1}$P$_{1}$ sector while,
in agreement with~\cite{vaec88}, our calculations reveal no state
with dominant 4d5p character clearly pointing to a strong departure
from an independent-particle description.

In the present work, our interest centers on the spectrum of high-$n$
strontium Rydberg states associated with excitation of a single electron.
While a few perturber states are important in the low $n$ regime,
the high-$n$ spectrum remains largely unaffected. The quantum numbers
$(n_{1},\ell_{1})$ of the outer Rydberg electron extend over a wide
range; however, those $(n_{2},\ell_{2})$ of the inner electron are
dominated by a limited number of low-lying states. We can thus truncate
the expansion in terms of orbitals of the inner electron to these
dominant orbitals which greatly facilitates the numerical diagonalization.
More precisely, all numerical results presented here employ six inner
electron $n\ell$ orbitals (5s, 4d, 5p, 6s, 5d, and 6p). For a fixed
component, $M=0$, and an excited state with principal quantum number
$n=50$, the total Hamiltonian has a dimension of about $40\,000$.
At vanishing electrical field, the latter consists of {\em uncoupled}
blocks of constant total angular momentum $L$. However, for excitation
in a dc electric field as considered below, the relevant energy in
the vicinity of the target level is identified and the resulting Stark
matrix to be diagonalized can have a dimension as low as $1\,000$,
rendering our truncated-basis approach numerically inexpensive. 
Convergence is checked by increasing the number of inner electron 
orbitals until the relative error in the eigenenergies 
compared with the results using 14 inner electron orbitals
is below 0.1\%. For high $n$ states below the first ionization 
threshold the present configuration-interaction
method should provide accuracies that are at least comparable to, if not even higher than, 
R-matrix methods which solve the eigenproblem within a reaction volume 
of finite radius $r_{0}$, the latter being determined by the largest 
extent of the inner electron configurations.

\subsection{Scaling}

Since quantum calculations of spectra and transition probabilities
for $n\gtrsim300$ using the present TAE model are not tractable,
we employ scaling relations to extrapolate our converged numerical
results for the range $10\lesssim n\lesssim50$ to very high $n$.
As a starting point, we employ the well-known classical scaling relations
for the Rydberg states in a pure Coulomb potential: \\
\begin{subequations} \label{eq:scaling} 
for energies, 
\begin{equation}
\frac{E_{n}}{E_{n_{0}}}=\left(\frac{n_{0}}{n}\right)^{2}\,,
\end{equation}
for electric field strength, 
\begin{equation}
\frac{F_{n}}{F_{n_{0}}}=\left(\frac{n_{0}}{n}\right)^{4}\,,
\end{equation}
for dipole moment, 
\begin{equation}
\frac{d_{n}}{d_{n_{0}}}=\left(\frac{n}{n_{0}}\right)^{2}\,,
\end{equation}
for transition matrix elements from low-lying states $i$ to Rydberg
states, 
\begin{equation}
\frac{d_{i,n}}{d_{i,n_{0}}}=\left(\frac{n_{0}}{n}\right)^{3/2}\,,
\end{equation}
and for the critical fields, $F_{{\rm cross},n}\simeq1/(3n^{5})$,
at which states in adjacent $n$ manifolds first cross 
\begin{equation}
\frac{F_{{\rm cross},n}}{F_{{\rm cross},n_{0}}}
=\left(\frac{n_{0}}{n}\right)^{5}\,.
\end{equation}
 \end{subequations} In the presence of the Sr$^{2+}$ core and the
additional valence electron, the scaling in $n$ {[}Eq.~(\ref{eq:scaling}){]}
can be extended to the effective quantum number 
\begin{equation}
n^{*}=n+\delta_{\ell}+\beta_{\ell}E_{n}\label{eq:ritz}
\end{equation}
 of the Rydberg-Ritz formula~\cite{jast48}. In Eq.~(\ref{eq:ritz}),
$\delta_{\ell}$ is the $n$ independent but $\ell$ dependent quantum
defect while the $n$ (or energy) dependent correction is given by
the Ritz coefficient, $\beta_{\ell}$. Our strategy is now to determine
$\beta_{\ell}$ and $\delta_{\ell}$ by diagonalizing the model Hamiltonian
Eq.~(\ref{eq:hamil}) in the basis {[}Eq.~(\ref{eq:basisRyd}){]}
truncated at $n=85$. From the resulting series of energy eigenvalues
and transition matrix elements ($n\gtrsim20$), we can, provided they
are free from the influence of perturber states, extrapolate to very
high $n$ employing the scaling relationships {[}Eq.~(\ref{eq:scaling}){]}
and the effective quantum numbers $n^{*}$.

\section{Zero-field photoexcitation}

\subsection{Quantum defects}

\begin{figure}[p]
\includegraphics[width=14cm]{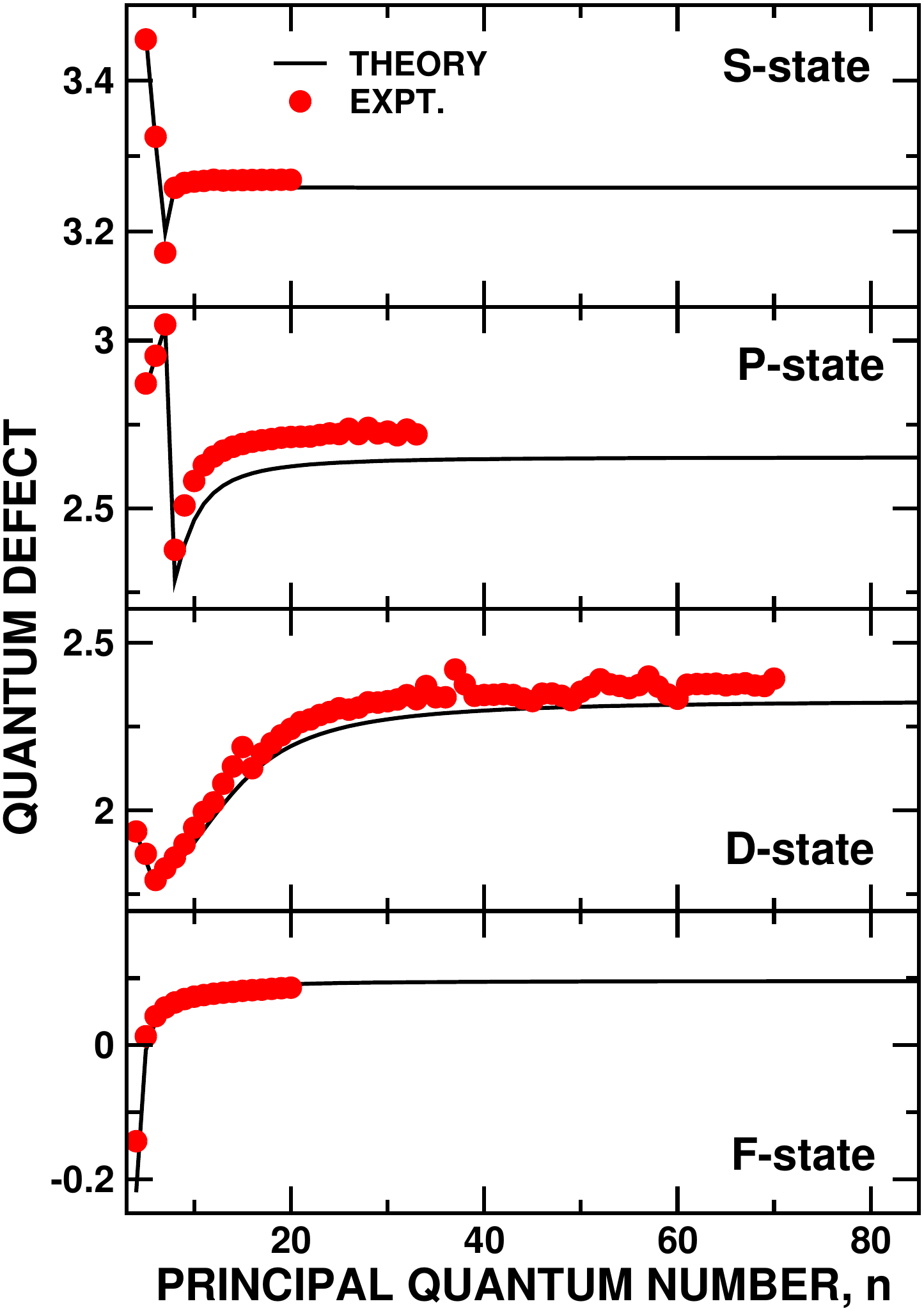}
\caption{ \label{fig:qd} (Color online) Measured and calculated quantum defects
for low-$n$ Rydberg states in strontium. The calculations (solid
lines) employ a two-active-electron model with six configurations (5s,
4d, 5p, 6s, 5d, and 6p) of the inner electron. Measured results (filled
circles) are taken from \cite{sans10} for $n\le20$, \cite{eshe77}
for P-states ($n>20$), and \cite{dai95} for D-states ($n>20$). }
\end{figure}

As a first test of the TAE model we compare (see Fig.~\ref{fig:qd})
the energies of low-$L$ and intermediate $n$ Rydberg states, expressed
in terms of $n$-dependent quantum defects, $\mu(n)=n-n^{*}$, 
[Eq.~(\ref{eq:ritz})] with experimental 
data~\cite{vail12,sans10,eshe77,dai95} . Overall,
our TAE energy values agree well with the experimental data 
as do the results of multichannel quantum defect theory~\cite{eshe77}
and of R-matrix calculations~\cite{ayma96} which employ the same semi-empirical
model potential. However, a small discrepancy between theory and experiment
is seen for the quantum defects of the P and D states, i.e., 
the calculated values slightly underestimate the measured quantum defects and we return to this in a moment. 
The effective quantum defects increase with $n$, especially for 
D states yielding a non-vanishing Ritz coefficient 
$\beta_{2}\simeq-43$. The value for P states, $\beta_{1} \simeq -8$,
is significantly smaller while the coefficient for S states 
$\beta_{0} \simeq 0.08$ is close to zero. 
In the series limit, the calculated values of the
$n$-independent quantum defects $\delta_{0}=3.26$, $\delta_{1}=2.65$
and $\delta_{2}=2.33$ are slightly lower than the measured values
$\delta_{0}=3.26$, $\delta_{1}=2.71$ and $\delta_{2}=2.38$. For
lower $n$, $\mu(n)$ deviates strongly from simple single-electron
Coulomb scaling signaling the presence of perturber states in the
two-electron system. These deviations are, however, of negligible
importance for the high-$n$ limit of interest here.
Adjusting the model-potential parameters and/or including dielectric 
core polarization terms~\cite{luck98} yields nearly identical results, still 
underestimating the measured quantum defects. This is due to the fact 
that, for all model potentials employed, the calculated energy 
levels of the Sr$^+$ ion exhibit deviations from the measured 
values which are of the same order of magnitude. The precise estimation
of the quantum defect, however, requires accurate values not only of the energy of the excited
state but also of the ionization threshold.
Thus, small but finite errors in ionic energies are passed on
to the energy values of neutral strontium and become non-negligible
for high-$n$ levels. Alternatively, it is possible to optimize 
the model potential in order to minimize these deviations.
For the present two-photon excitation study we chose the
model potential of Ref.~\cite{ayma96} since the intermediate 5s5p $^1$P$_1$
state as well as the excited states in the S- and D-sectors are reasonably
well represented (see Fig.~\ref{fig:qd}). 

\subsection{Oscillator strengths}

\begin{figure}
\includegraphics[width=14cm]{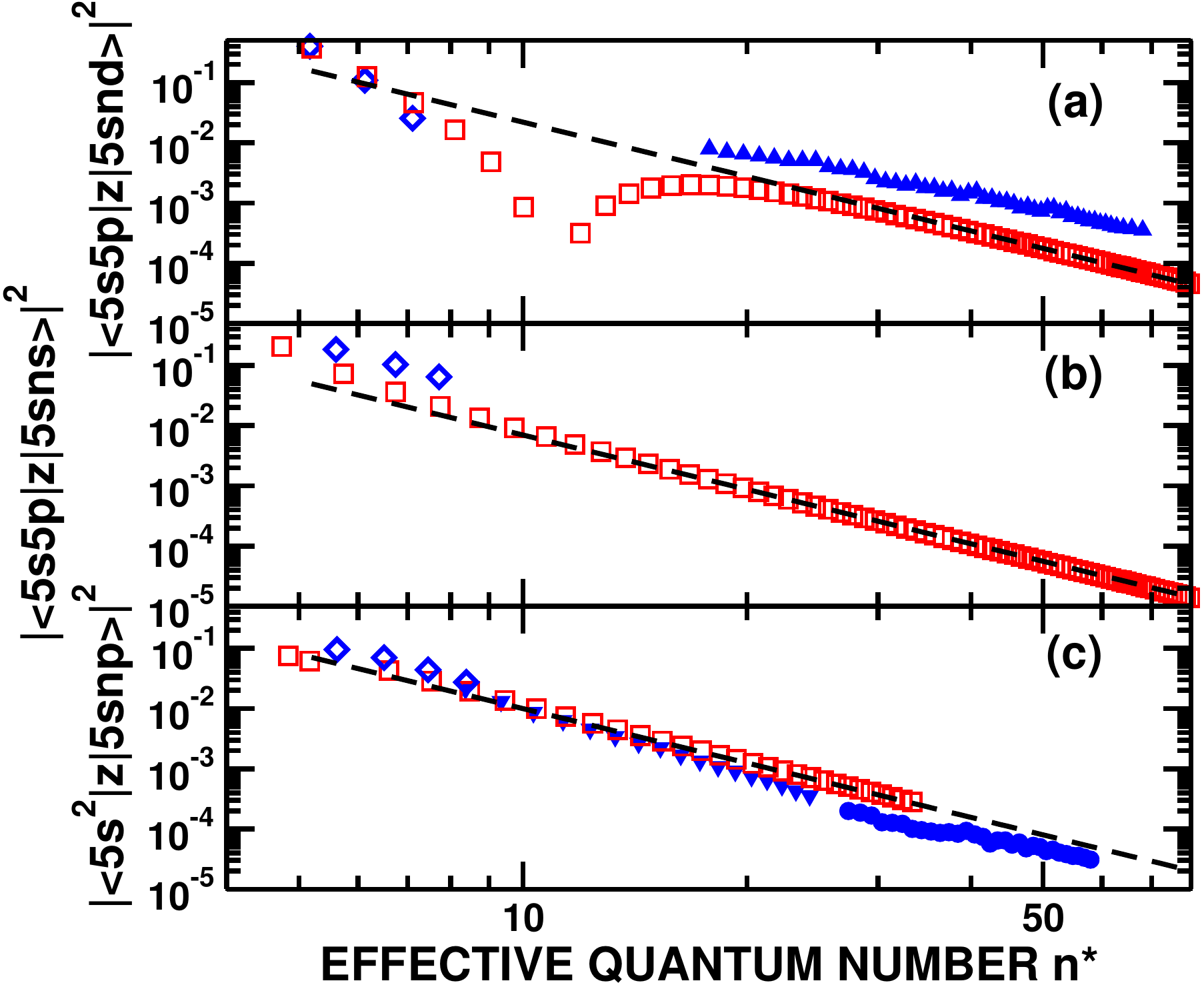}
\caption{\label{fig:scaling_dipo} (Color online) Absolute magnitude squared
of the calculated dipole matrix elements (a) $|\langle{\rm 5s5p}|z|5{\rm s}n{\rm d}\rangle|^{2}$,
(b) $|\langle{\rm 5s5p}|z|{\rm 5s}n{\rm s}\rangle|^{2}$ and (c) $|\langle{\rm 5s}^{2}|z|{\rm 5s}n{\rm p}\rangle|^{2}$
as a function of the effective principal quantum number $n^{*}$.
($\Box$ : calculations employing a TAE model with
the six inner electron orbitals
as in Fig.~\ref{fig:qd}, $\diamond$ : calculations from \cite{weri92},
$\blacktriangle$ : measured data~\cite{haq07}, $\bullet$
: measured data~\cite{mend97}, and $\blacktriangledown$ : measured
data~\cite{gart83}.) The initial dip in (a) around $n^{*}=10$ ($n=13$)
represents a Cooper minimum followed by a smooth approach to the expected
$n^{*\,-3}$ dependence (dashed black is drawn line to guide the eye.) }
\end{figure}

A second test of the TAE model is provided by comparing the calculated
oscillator strengths with those measured experimentally 
(see Fig.~\ref{fig:scaling_dipo})
both for the 5s$^{2}$ $^{1}$S$_{0}$ $\to$ 5s5p $^{1}$P$_{1}$
transition, and for the subsequent transitions to the Rydberg manifolds,
5s5p $^{1}$P$_{1}$ $\to$ 5s$n$s $^{1}$S$_{0}$ and 5s$n$d $^{1}$D$_{2}$.
The first transition provides a sensitive test of the limitations
of independent-electron models. This transition is strongly influenced
by configuration mixing, closely connected to the inclusion of the
4d5p $^{1}$P$_{1}$ configuration among the members of the 5s$n$p
$^{1}$P$_{1}$ manifold mentioned previously. The collective, i.e.,
multi-electron, character of the transition is reflected in an oscillator
strength exceeding unity. The value from our TAE model $f_{{\rm 5s5p}}=2.02$
agrees with the experimental value $f_{{\rm 5s5p}}^{{\rm expt}}=1.91$
and is in reasonable agreement with the calculations by Vaeck et al.
\cite{vaec88} and Werij et al. \cite{weri92}. The subsequent 5s5p
$^{1}$P$_{1}$ $\to$ 5s$n$s $^{1}$S$_{0}$ or 5s$n$d $^{1}$D$_{2}$
transitions control the excitation of the Rydberg manifolds of interest
here and allow for a test of $n$-scaling of the oscillator strengths
and transition probabilities. For low $n$ ($n\le20$) significant
deviations from simple Coulomb $n$ scaling [Eq.~(\ref{eq:scaling}c)] are observed. Most prominent
is the Cooper minimum near $n=13 (n^{*}=10)$ in the excitation probability
to the $^{1}$D$_{2}$ states. The observed minimum is associated
with a change of sign in the dipole matrix elements. To our knowledge,
this Cooper minimum has not been reported in the literature. We note
that the vanishing dipole coupling in the vicinity of the Cooper minimum
increases the influence of otherwise weak effects, such as spin-orbit
interactions. The latter are, of course, not captured by our non-relativistic
Hamiltonian {[}Eq.~(\ref{eq:hamil}){]}. Experimental data~\cite{eshe77}
indicate that at $n\approx15$ the singlet and triplet $L=2$ Rydberg
series of strontium undergo an avoided crossing, i.e., $L$-$S$ coupling
breaks down. Our calculations agree well with the calculations by
Werij et al~\cite{weri92} for $n^{*}<8$ but underestimate the spectroscopic
data for higher $n$ ($20\le n\le68$)~\cite{haq07}. The oscillator
strength smoothly approaches the $(n^{*})^{-3}$ scaling from below.
To further check the validity of our calculations, the dipole matrix
elements for $\langle{\rm 5s5p}^{1}{\rm P}_{1}|z
|{\rm 5s}n{\rm s}^{1}{\rm S}_{0}\rangle$
(Fig.~\ref{fig:scaling_dipo}b) and 
$\langle{\rm 5s}^{2}\,^{1}{\rm S}_{0}|z|{\rm 5s}n{\rm p}^{1}{\rm P}_{1}\rangle$
(Fig.~\ref{fig:scaling_dipo}c) transitions are compared with available
data. Our calculations agree reasonably well with the results of 
earlier calculations~\cite{weri92} for the former 
(Fig.~\ref{fig:scaling_dipo}b),
and the measurements~\cite{gart83,mend97} for the latter 
(Fig.~\ref{fig:scaling_dipo}c).
We note that the dipole matrix elements are evaluated from the oscillator
strength, $f$, or the transition probability, $A$, published
in the references mentioned above assuming the relationships
$f = 2 \omega | \langle 5{\rm s}nl | z | 5{\rm s}n'l' \rangle |^2$
and $A = 4 \omega^3 | \langle 5{\rm s}nl | z | 5{\rm s}n'l' \rangle |^2/
(3 c^3)$.

\begin{figure}
\includegraphics[width=14cm]{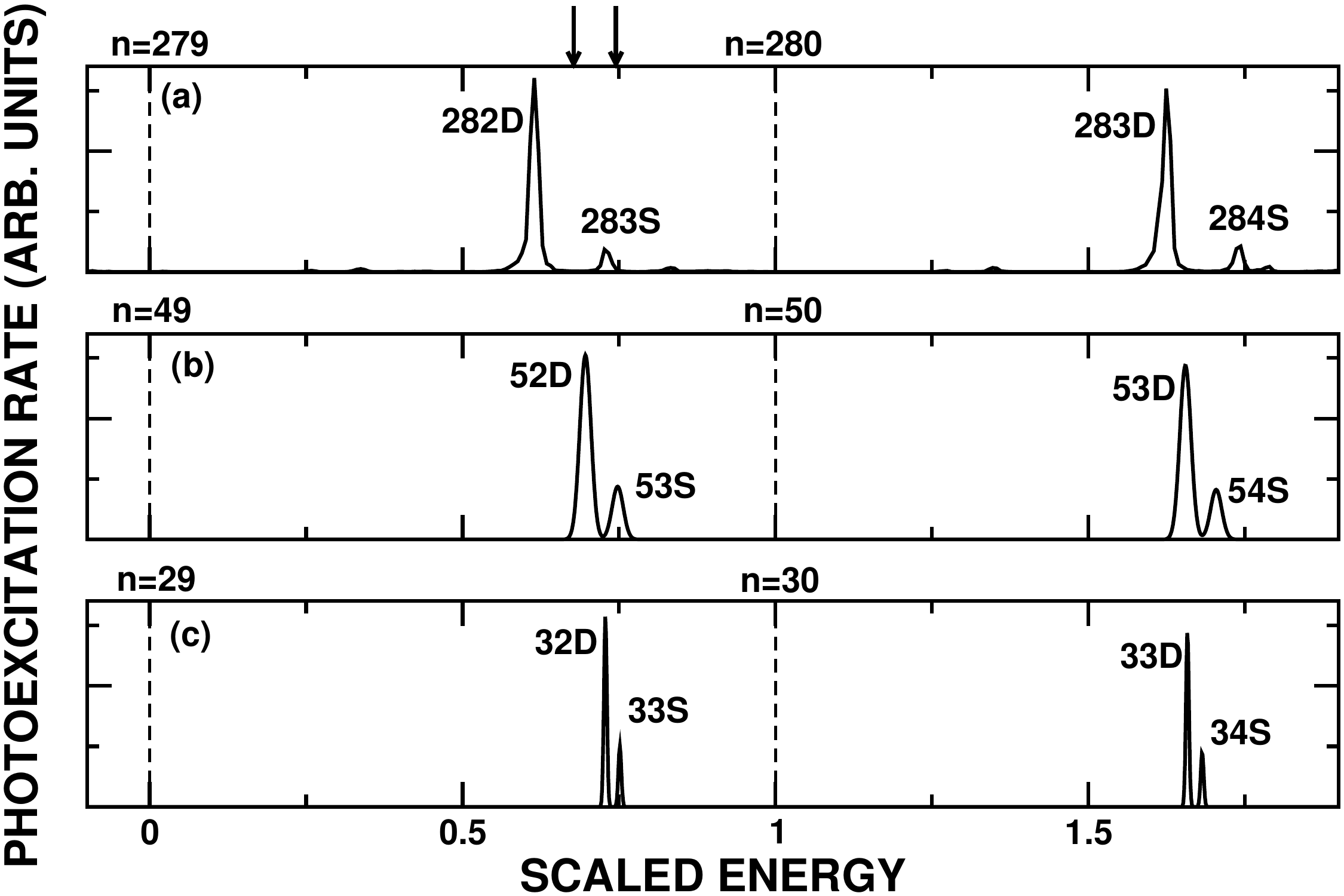}
\caption{\label{fig:zeroF} Comparison between measured and
calculated (zero-field) excitation spectra. (a) Measured excitation
spectrum recorded at $n\sim283$. (b), (c) Results of two-electron
calculations at $n\sim50$ and $n\sim30$, respectively, employing
the same six inner electron states as in Fig.~\ref{fig:qd}. The
arrows indicate the energy for the 282d and the 283s states predicted
by the scaling relations {[}Eqs.~(\ref{eq:scaling},\ref{eq:ritz}){]}
when $\delta_{\ell}$ and $\beta_{\ell}$ are fitted to the numerically
calculated spectrum of the TAE model for $25\le n\le85$. To facilitate
the comparison between different $n$, we exploit the scaling relations
{[}Eq.~(\ref{eq:scaling}){]}, a scaled energy of one corresponding
to the energy difference between adjacent $n$ and $(n-1)$ manifolds.}
\end{figure}

In the following we estimate the production rates for 5s$n$d $^{1}$D$_{2}$
and 5s$n$s $^{1}$S$_{0}$ Rydberg atoms near $n\simeq300$. 
Fig.~\ref{fig:zeroF}
illustrates the evolution of the photoexcitation spectrum from $n=39$
to 280. The calculated spectra for $n=30$ and 50 are compared with
spectra measured near $n=280$. The calculated spectrum for $n=50$
is convoluted with a Gaussian to match the (scaled) measured experimental
line width while for $n=30$ a smaller line width is chosen to resolve
the S- and D-states (near $n\simeq30$ the spacing of the $(n+1)$S and
$n$D features is very small). The width of the measured lines is
principally attributed to transit time broadening (the transit time
of an atom through the 413~nm laser spot is $\sim$~300~ns) with
smaller contributions from the finite laser pulse width and fluctuations
in the laser frequencies during the $\sim$~1~s accumulation time
for each data point in the spectrum. The energy axes in Fig.~\ref{fig:zeroF}
are scaled as in Eq.~(\ref{eq:scaling}). The arrows in Fig.~\ref{fig:zeroF}a
show the predictions for the position of the spectral lines based
on the Rydberg-Ritz scaling {[}Eq.~(\ref{eq:ritz}){]} derived from
the calculated spectrum for lower $n$. While the position of the
S-state is well predicted, the D-state position is slightly off due
to the underestimated quantum defect {[}Fig.~\ref{fig:qd}{]}. The
positions of the $n$ $^{1}$S$_{0}$ states relative to the two adjacent
$n$ manifolds are nearly invariant with respect to $n$ as the quantum
defects of the $n$$^{1}$S$_{0}$ states are nearly $n$-independent.
On the other hand, the relative positions of the $n$ $^{1}$D$_{2}$
states vary with $n$ mirroring the $n$-dependent quantum defect.
The measured relative size of the $^{1}$D$_{2}$ to $^{1}$S$_{0}$
lines near $n=280$ is $\sim8.7$. For lower $n$ the calculations
indicate a moderate increase of the $^{1}$D$_{2}$$/$$^{1}$S$_{0}$
ratio and -- in line with the $n^{*}$ scaling -- predict a ratio of $\sim 4$
at $n = 280$. The origin of this discrepancy is not yet clear. 
Experimentally, stray fields might
enhance this ratio, as might any deviations from collinear alignment
of the polarization axes of the two lasers. 
As will be discussed below, the measured excitation rate of 
$^{1}$D$_{2}$ states is smaller than the theoretical prediction
indicating that the enhanced ratio is due to a suppression of
$^{1}$S$_{0}$ state excitation rather than an enhancement of
$^{1}$D$_{2}$ state excitation.

It is instructive to compare the excitation spectrum calculated within
the present TAE approximation with the results of a single-active-electron
(SAE) model with a model potential similar to that described 
in Ref.~\cite{mill11}.
We have found that the quantum defects resulting from the SAE model
display a very weak $n$-dependence. That is, the eigenenergies and
the quantum defects can be obtained quite accurately for an optimized
model potential but only for a limited range of principal quantum
numbers. When adjusting the model parameters to fit the asymptotic
quantum defects, the high Rydberg states are reasonably well represented
by the SAE model. However, the calculated oscillator strengths fail
to reproduce the measured excitation spectra which we trace to an
inaccurate description of the 5s5p state. This failure reflects the
fact that the first excitation step is strongly influenced by electron-electron
interactions not accounted for by the SAE model.

\subsection{Rydberg excitation rates}

For applications involving strongly coupled Rydberg atoms, it is
crucial to achieve high excitation densities which poses a challenge given the $(n^{*})^{-3}$
scaling of the oscillator strength. We have therefore measured the
strontium Rydberg excitation rates that can be achieved in the beam.
Measurements at $n\sim310$ show that the probability to create
 a Rydberg atom at the peak of a $^{1}$D$_{2}$ feature during
a $\sim$~500~ns laser pulse can be made large. To avoid saturation effects, 
in the standard setup
the probability is purposely maintained below $\sim$~0.4 by substantially
reducing the strontium atom beam density by operating the oven at
$\sim$~500~$^{\circ}$C. However, tests undertaken at higher operating
temperatures in which the 413~nm laser beam was attenuated using
neutral density filters to limit the excitation rate showed that,
with the oven operating at $\sim$~630$^{\circ}$C and using the
full 413~nm laser power ($\sim$70~mW), $\sim$15 Rydberg atoms
can be produced per laser pulse which, when allowance is made for
the motion of atoms in the beam during the 500~ns-duration laser
pulse, corresponds to a product Rydberg atom density of 
$\sim3\times10^{5}$~cm$^{-3}$.

To a good first approximation, the excitation rate can be estimated
by assuming two sequential one-photon processes. The first transition
(Rabi period $0.1\mu$s) may be considered as saturated having a Lorentzian
profile with a width of $\sim$~10MHz. For an intensity of 250~W~cm$^{-2}$,
the second transition has a width $\omega\sim0.5$~MHz and thus a
Rabi period of about 2$\mu s$. Given the relatively long atomic transit
time, $t_{{\rm transit}}\sim300$~ns, through the 413~nm laser beam,
a linear (Fermi Golden Rule) estimate would likely overestimate the
excitation probability. Therefore, this probability is obtained using
the Rabi formula $P(\Delta)=\omega^{2}(1-\cos[(\sqrt{\omega^{2}+\Delta^{2}})t_{{\rm transit}}])/2(\omega^{2}+\Delta^{2})$,
where $\Delta$ is the detuning. If the excitation is considered as an
incoherent sequence of two transitions, the total Rydberg excitation
probability is given by the integral (over the detuning) of the product
of the two Lorentzian excitation profiles. This approach yields an
excitation probability of $\sim$~1\% per atom. Given the estimated
strontium atom beam density of $\sim10^{8}$~cm$^{-3}$ this would
suggest creation of Rydberg atom densities of $\sim10^{6}$~cm$^{-3}$ Rydberg atoms per laser pulse which
are somewhat higher than the experimental measurements. 

Rydberg atom densities of $\sim3\times10^{5}$~cm$^{-3}$ result in a typical
inter-Rydberg spacing of $\sim$150~$\mu$m which is approaching
distances at which effects due to Rydberg-Rydberg interactions such as
blockade become important. Blockade results when the energy level
shifts induced by the excitation of a Rydberg atom prevent the excitation
of other neighboring atoms. Calculations of such shifts have been
undertaken for strontium $n{}^{1}$D$_{2}$ (and other) states for
values of $n$ up to $\sim$70~\cite{vail12}. Extrapolation of these
results to $n\sim300$ would lead to blockade radii of $\sim$500~$\mu$m
for an effective laser line width of $\sim$5~MHz. The validity of
this extrapolation to very high $n$ is, however, an open question.
An alternative estimate for the blockade radius can be obtained by
considering the dipole-dipole interaction of two permanent dipole
moments $\vec{d}_{1}$, $\vec{d}_{2}$, $\propto d_{1}d_{2}/R^{3}$,
which for extreme high-$n$ parabolic states are of the order of $\sim n^{2}$~a.u.
This model suggests a blockade radius of $\sim150\,\mu$m at $n\sim300$.
As an aside we note that the number of Rydberg atoms produced in the
present experiments with strontium is much higher than achieved
for potassium at comparable beam densities in earlier experiments~\cite{dunn09}.
This increase is primarily due to a threefold increase in the laser
powers available and a factor of three decrease in the effective experimental
line width.

\subsection{Isotope shifts }

\begin{figure}
\includegraphics[width=9cm]{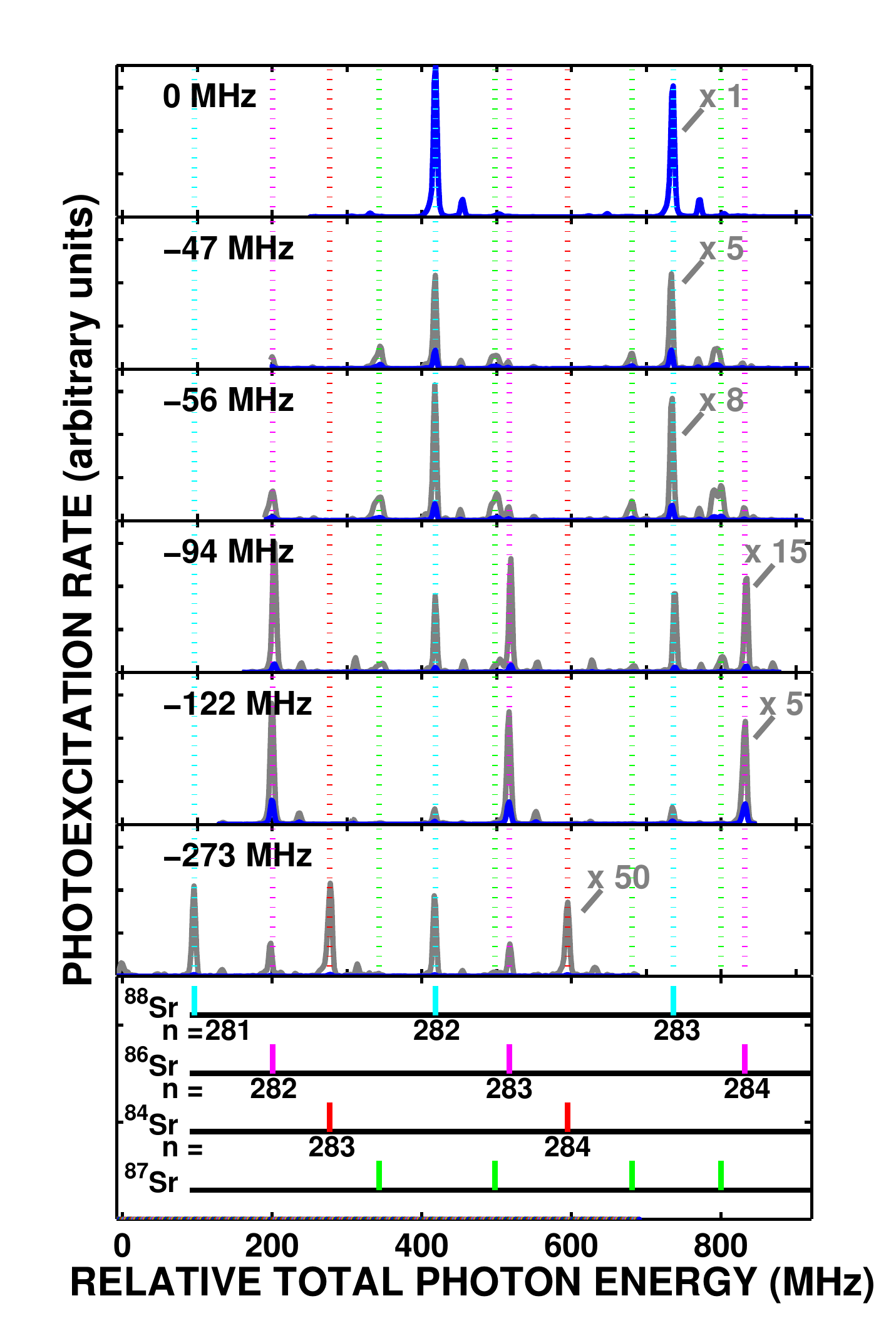}
\caption{\label{fig:isotope} (Color online) Excitation spectra recorded in
the vicinity of $n=283$ for different detunings of the 461~nm laser.
These detunings relative to the $^{88}$Sr 5s$^{2}$ $^{1}$S$_{0}$
- 5s5p $^{1}$P$_{1}$ transition, are indicated (see also Table.~I).
Specifically, a detuning of $\sim-122(-273)$~MHz optimizes the transition
in the $^{86}$Sr ($^{84}$Sr) isotope, while detunings of about $-47$
and $-56$~MHz are chosen to favor excitation of $^{87}$Sr. The
frequency axis shows the sum of the 461~nm and 413~nm photon energies.
The horizontal bars beneath the data identify the position of the
5s$n$d $^{1}$D$_{2}$ Rydberg states in the $^{88}$Sr, $^{86}$Sr,
and $^{84}$Sr isotopes, and of features attributed to excitation
of $^{87}$Sr Rydberg states. }
\end{figure}

As a test of the spectral resolution achievable in the present experimental
apparatus we probe the isotope shifts in very high-$n$ Rydberg atoms
near the series limit. Fig.~\ref{fig:isotope} shows zero-field
excitation spectra recorded in the vicinity of $n=282$ for various
detunings of the 461~nm laser which were selected to optimize the
5s$^{2}$ $^{1}$S$_{0}$ $\to$ 5s5p $^{1}$P$_{1}$ transition in
the different strontium isotopes. The relative frequencies of these
transitions together with other properties of naturally-occurring
strontium are listed in Table I~\cite{beig82,elie83,maug07}. In
both Table I and Fig.~\ref{fig:isotope}, the isotope shifts and
hyperfine splittings are quoted relative to the most abundant $^{88}$Sr
isotope. The frequency axis in Fig.~\ref{fig:isotope} shows the
sum of the 461~nm and 413~nm photon energies.

The reference spectrum obtained with the 461~nm laser tuned to optimally
excite the dominant $^{88}$Sr isotope (upper panel) displays a series
of sharp peaks associated with the excitation of $^{1}$D$_{2}$ and
$^{1}$S$_{0}$ states (see the spectrum of Fig.~\ref{fig:zeroF}a).
As the 461~nm laser is red detuned from resonance with the $^{1}$S$_{0}$
$\to$ $^{1}$P$_{1}$ transition in $^{88}$Sr (lower panels in Fig.~\ref{fig:isotope}),
the size of the $^{88}$Sr features in the excitation spectra decrease
steadily and new features associated with the excitation of Rydberg
states of the other isotopes emerge. Rather easy to identify are the
spectral features seen at a detuning of $\sim-122$~MHz, which optimizes
the $^{1}$S$_{0}$ $\to$ $^{1}$P$_{1}$ transition in the $^{86}$Sr
isotope. The excitation spectrum is dominated by the creation of $^{86}$Sr
$^{1}$D$_{2}$ states although some residual excitation of $^{88}$Sr
isotopes remains. The observed $^{88}$Sr-$^{86}$Sr isotope shift
in the series limit, $\sim+210\pm5$~MHz, is in agreement with the
value obtained by extrapolation of earlier spectroscopic studies at
lower $n$~\cite{lore83}. Similarly, at a blue laser detuning of
$\sim-273$~MHz which optimizes the $^{1}$S$_{0}$ $\to$ $^{1}$P$_{1}$
transition in the $^{84}$Sr isotope, the excitation of $^{84}$Sr
$^{1}$D$_{2}$ Rydberg states becomes apparent. The observed $^{88}$Sr-$^{84}$Sr
isotope shift in the series limit, $+440\pm8$~MHz, is again consistent
with earlier measurements at lower $n$~\cite{lore83}. The fractional
abundances of the $^{86}$Sr and $^{84}$Sr isotopes in the beam (see
Table I) are directly reflected in the relative intensities of the
excitation peaks.

Several features are observed in the excitation spectra at blue laser
detunings chosen to favor excitation of the $^{87}$Sr isotope (at
detunings of about $-47$, and $-56$~MHz). This (odd) isotope displays
strong hyperfine-induced singlet-triplet mixing and strong interactions
between states of different $n$~\cite{beig88,sun89}. A detailed
analysis of these features requires stray fields to be reduced to very
low levels~\cite{beig88} well beyond the capabilities of the present
apparatus.

\subsection{Extension to higher $n$}

\begin{figure}
\includegraphics[width=12cm]{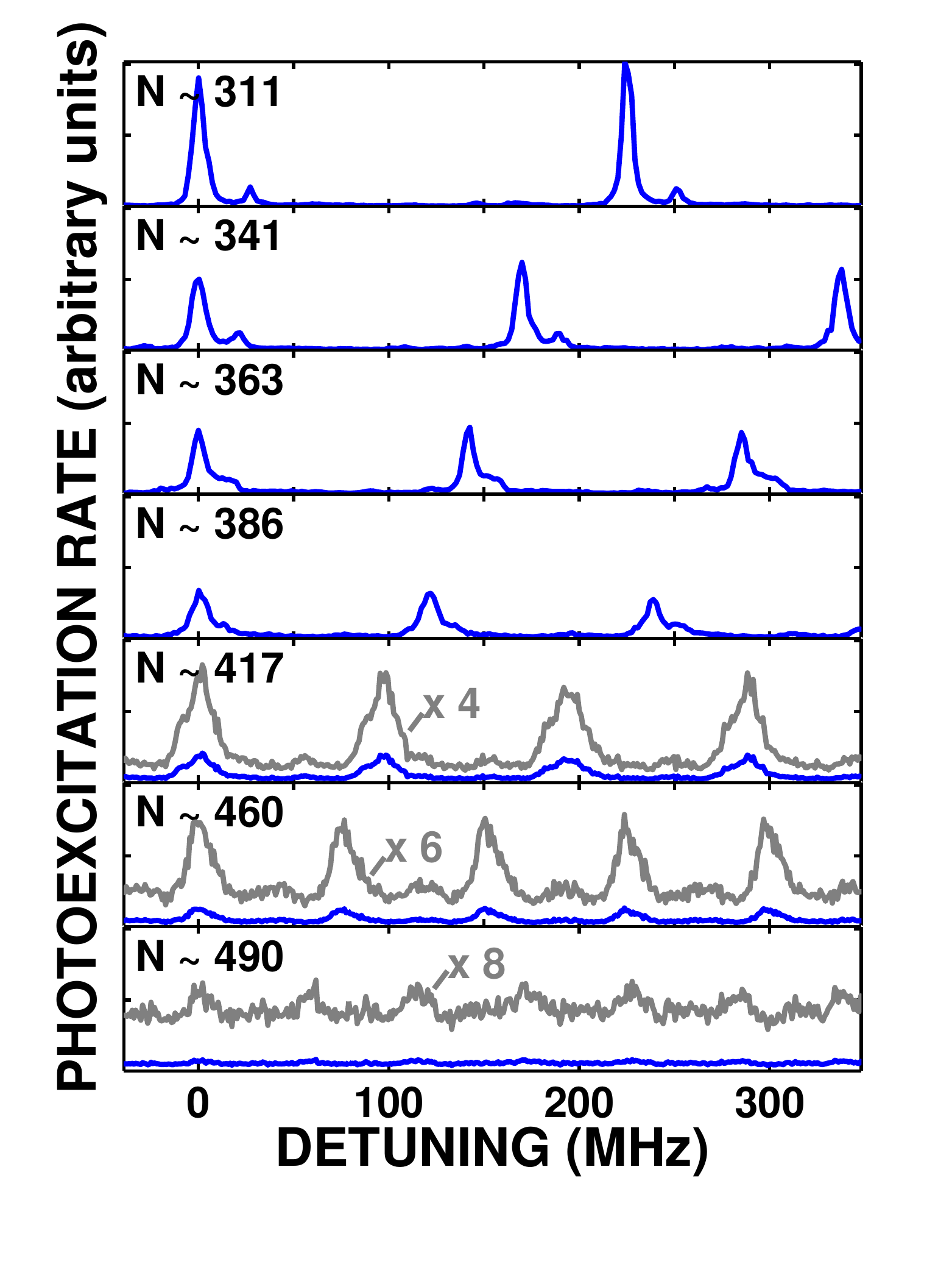}
\caption{ \label{fig:high_n} (Color online) Excitation spectra recorded near
the values of $n$ indicated. The 461~nm laser is tuned to the $^{88}$Sr
5s$^{2}$ $^{1}$S$_{0}$ - 5s5p $^{1}$P$_{1}$ transition. The frequency
axis shows the relative frequency of the 413~nm laser during each
scan. }
\end{figure}

We close the discussion of zero-field excitation with Rydberg spectra
recorded at even higher values of $n$ and the 461~nm laser tuned
again to the $^{1}$S$_{0}$ $\to$ $^{1}$P$_{1}$ transition in
the dominant $^{88}$Sr isotope. It is evident from Fig.~\ref{fig:high_n}
that, as $n$ increases, the number of Rydberg atoms created decreases
dramatically as a result of both the decrease in the oscillator strength
and the increasing width of the spectral features. For values of $n\le350$,
two well-resolved Rydberg series are seen, corresponding to excitation
of $^{1}$D$_{2}$ and $^{1}$S$_{0}$ states. With further increases
in $n$ the spectral features begin to broaden significantly, their
widths having approximately doubled by $n\sim400$. For even larger
values of $n$ the background signal begins to increase, but a well-resolved
Rydberg series is still evident for values of $n$ up to $\sim$460.
For $n>500$, however, it becomes increasingly difficult to discern
any Rydberg series. This degradation in the Rydberg spectrum in the
limit of very high $n$ can be attributed to the presence of very
small stray background fields in the excitation volume. As discussed
below, the effects of such fields become particularly important when
they approach the fields at which states in adjacent Stark manifolds
cross, $\sim$ 50~$\mu$V~cm$^{-1}$ at $n\sim500$. This suggests
that stray fields of $\sim$ 50~$\mu$V~cm$^{-1}$ remain in the
excitation volume which is consistent with earlier estimates of their
size.

\section{ Photoexcitation in a dc field }

We focus now on the Stark spectrum for strontium. More specifically,
we investigate the spectrum of the singly-excited Rydberg states as
a function of the strength $F_{{\rm dc}}$ of a dc field applied along
the $z$ axis. Well-controlled excitation of Stark states is key to
creation of ensembles of Rydberg atoms with large permanent dipole
moments.

\begin{figure}
\includegraphics[width=15cm]{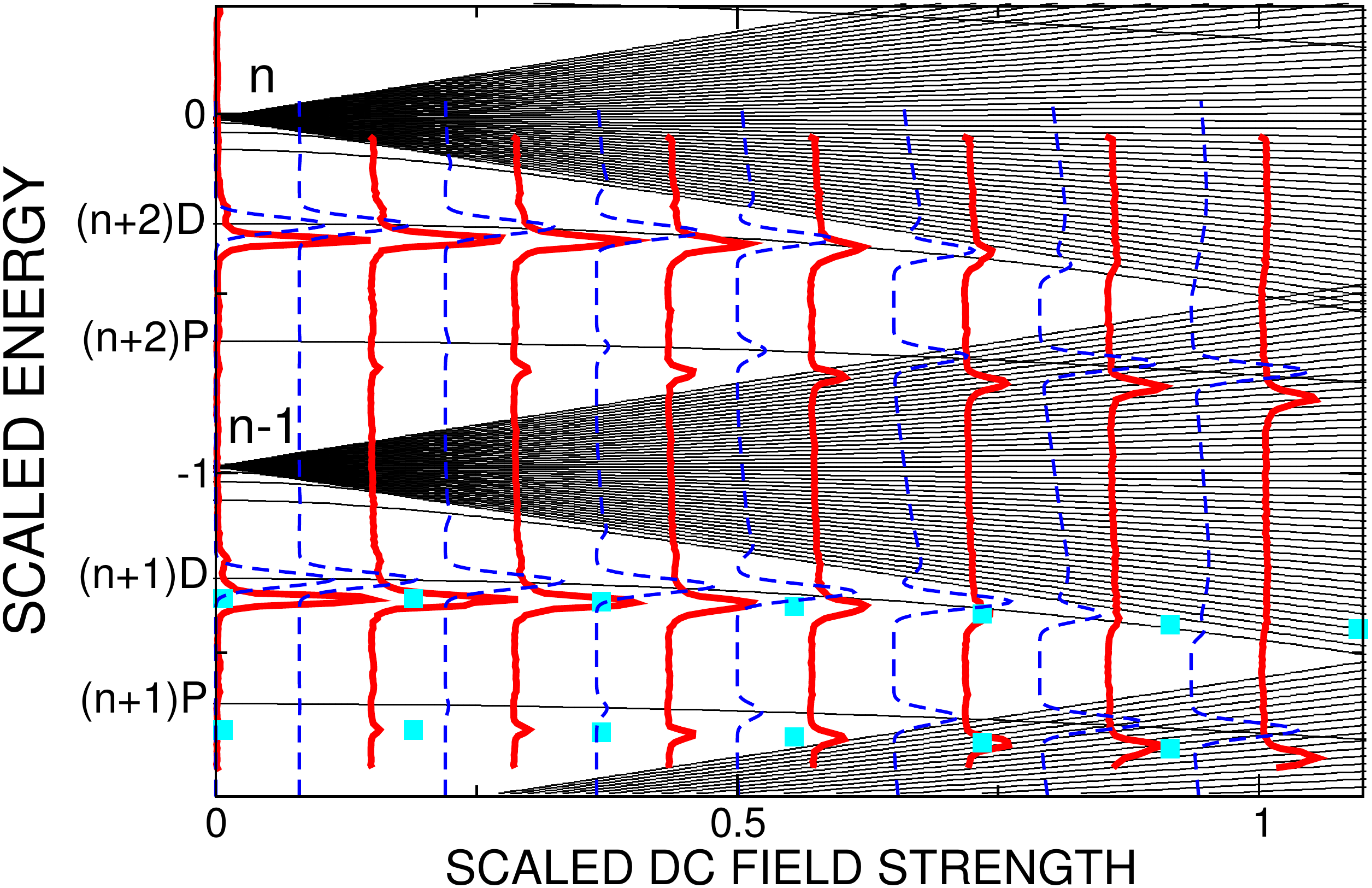} 
\caption{\label{fig:stark} (Color online) Evolution of the experimental excitation
spectrum with increasing dc applied field for M=$\pm$1 states in
the vicinity of $n\sim310$ (thick red line) together with the calculated
eigenenergies (thin solid black lines) and the excitation spectrum
(dashed blue lines) of singly-excited strontium for $n\simeq50$.
The squares denote the energies of earlier measurements at $n\sim80$~\cite{mill10}.
Data for the very different $n$ levels are compared by employing
scaled field and energy axes. }
\end{figure}

In contrast to the preceding discussion of excitation at vanishing
field, we consider orthogonal polarizations of the lasers (experimentally,
a half-wave plate was used to rotate the plane of polarization of
the 413~nm laser by $\sim90$$^{\circ}$). This leads to excitation
of Rydberg states with $M=\pm1$ and thus suppresses excitation of
$^{1}$S$_{0}$ states which, in turn, simplifies the excitation spectrum.
The calculated Stark spectrum near $n=50$ (Fig.~\ref{fig:stark})
illustrates the evolution of the excitation spectrum from the low-field
to the high-field regime characterized by the crossing of adjacent
manifolds. High-$L$ states, which are nearly degenerate at $F_{{\rm dc}}=0$,
exhibit linear Stark shifts. States from adjacent Stark manifolds
first cross at a field given by Eq.~(\ref{eq:scaling}e). Initially
the low angular momentum $^{1}$P$_{1}$ and $^{1}$D$_{2}$ states
exhibit only a quadratic Stark effect, a consequence of strong core
scattering which prevents the state from acquiring a large dipole
moment and limits its polarizability~\cite{heze92a}. Exploiting
the scaling relations, we display in Fig.~\ref{fig:stark} the theoretical
excitation spectra for field-perturbed $n$$^{1}$D$_{2}$ and $n$$^{1}$P$_{1}$
states near $n=50$ together with the experimental spectra near $n\simeq310$.
Fig.~\ref{fig:stark} also includes the results of previous measurements
at $n=80$~\cite{mill10}. In zero field ($F=0$) only $n$$^{1}$D$_{2}$
states can be excited from the intermediate 5s5p $^{1}$P$_{1}$ state.
As the strength of the dc field grows, the $n$$^{1}$D$_{2}$ states
increasingly couple with other angular momentum states. This increase
in $L$-mixing results in a marked decrease in the oscillator strengths
associated with their excitation. As each ``$n$$^{1}$D$_{2}$''
state merges with the linear Stark manifold, $L$-mixing is so strong
that the effects of core scattering become negligible and the state
becomes almost indistinguishable from the nearby extreme (strongly-polarized)
Stark states. The small difference in the scaled Stark spectrum
between the energy levels calculated (for the ``52$^{1}$D$_{2}$'') state 
and those measured (for the ``312$^{1}$D$_{2}$'') state result 
from the fact that the calculated zero-field quantum
defects are somewhat smaller than the experimental values (Fig.~\ref{fig:qd}).
These minor quantitative differences aside, the calculated and measured
excitation spectra are found to be in good accord. 

\begin{figure}
\includegraphics[width=16cm]{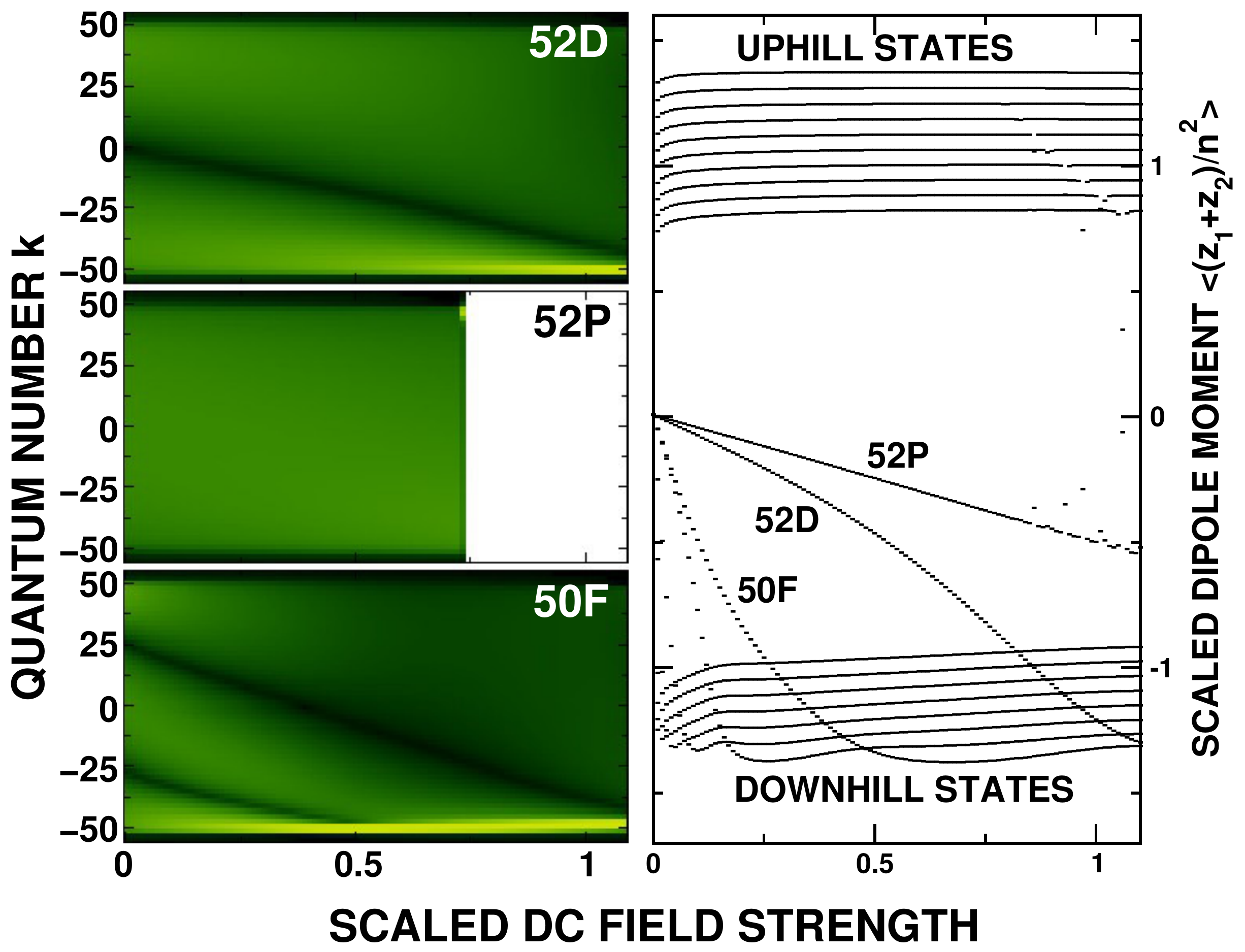}
\caption{\label{fig:dist_k} (Color online) Probability distribution of the
parabolic quantum number $k$ ($=-nA_{z}$) for the ``52D'', ``52P'',
and ``50F'' states (labeled by their zero-field-designation as a
function of the applied dc field $F_{{\rm dc}}$ normalized to the
crossing field $F_{{\rm cross}}$ (see text). The distribution for
the ``52P'' state is truncated where it merges with a Stark manifold.
The right hand figure shows the field dependence of the average dipole
moment for the same states as well as for representative extreme downhill
and uphill Stark states.\\[5cm]}

\end{figure}

In view of the
close correspondence between measurement near $n=310$ and theory
near $n=50$, we detail the evolution of the ``$n$$^{1}$D$_{2}$''
states with increasing dc field based on the computed energy spectrum
at $n\simeq50$. This aims at characterizing the transition from an
unpolarized $n^{1}D_{2}$ state to a strongly polarized Stark state.
This evolution can be visualized by considering the distribution of
projections onto states with parabolic quantum number $k$ 
\begin{equation}
\rho(k)=\sum_{n}\left|\,^{{\rm H}}\langle n,k,m|n_{{\rm Stark}}\rangle^{{\rm Sr}}\right|^{2}\,,\label{eq:dist_k}
\end{equation}
 or, equivalently, the distribution of the $z$-component, $A_{z}$,
of their Runge-Lenz vectors since $k$ corresponds to the quantized
action of $-nA_{z}$. Here $|n,k,m\rangle^{{\rm H}}$ are the hydrogenic
parabolic states and $|n_{{\rm Stark}}\rangle$ describes the numerically
exact eigenstate of the singly-excited strontium atom in the dc field.
Since the inner valence electron is almost exclusively in the 5s state,
i.e., $|n_{{\rm Stark}}\rangle$ is close to a natural orbital of
the two-electron system, the outer-electron state can be factorized
to evaluate the overlap. Fig.~\ref{fig:dist_k} displays the evolution
of the $k$-distribution for the ``52D'' $M=1$ state as a function
of applied field. For weak fields, the $k$-distribution is broad
covering a wide range of values between $-n$ and $n$ indicating
that the state is unpolarized. This can be easily understood by recalling
that the classical Runge-Lenz vector gives the direction of the major
axis of the Kepler ellipse. A wide distribution of $A_{z}$ implies
an ensemble of Kepler ellipses with $L=\ell+1/2$ whose major axes
are broadly distributed. For non-hydrogenic atoms that possess a sizable
quantum defect, such a distribution results from core scattering which
rotates the orientation of the ellipse while maintaining its eccentricity.
A minimum in the $k$-distribution is seen near $k=0$ which mirrors
a node of the spherical harmonic $Y_{\ell=2}^{m=1}$. As $F_{{\rm dc}}$
is increased this node shifts towards negative $k$ and the $k$-distribution
becomes increasingly asymmetric indicating that the state becomes,
indeed, gradually polarized. When in the Stark map (see Fig.~\ref{fig:stark})
the state merges with the neighboring Stark manifold, its $k$-distribution
becomes narrow consistent with evolution towards a narrow range of
parabolic states. This is in sharp contrast to the $k$-distribution
of the ``52P'' state {[}Fig.~\ref{fig:dist_k}{]} which shows little
sign of asymmetry even as the neighboring manifold is approached suggesting
that the state is only weakly polarized. The pronounced difference
between D and P states arises because P states couple strongly only
with adjacent S and D states, both of which are difficult to polarize.
The D states, on the other hand, directly couple with neighboring
F states which merge with the Stark manifold at relatively small values
of $F_{{\rm dc}}$ {[}see Fig.~\ref{fig:dist_k}{]} and become polarized.
The sizable polarization of the ``52D'' state, therefore, results
from strong coupling to this polarized ``50F'' state. In agreement
with this line of reasoning, we have found that the $M=0$ S state
is much less polarized than the $M=0$ D state.

We now quantify the degree of polarization in terms of dipole moments,
$\langle z_{1}+z_{2}\rangle$ {[}see Fig.~\ref{fig:dist_k}{]}. High-$L$
states, nearly degenerate at $F_{{\rm dc}}=0$, become polarized even
in very weak fields with dipole moments approaching the hydrogenic
field-independent value $\langle(z_{1}+z_{2})\rangle=(3/2)nk$. Well-isolated
low-$L$ states display a linear response behavior 
\begin{equation}
\langle z_{1}+z_{2}\rangle=-\alpha F_{{\rm dc}}\,\label{eq:avez}
\end{equation}
 with a quadratic Stark shift $\Delta E=-(1/2)\alpha F_{{\rm dc}}^{2}$.
To lower-order perturbation theory, the atomic polarizability $\alpha$
is given by 
\begin{equation}
\alpha=2\sum_{n'}\sum_{L'=L\pm1}\frac{|\langle nLM|(z_{1}+z_{2})|n'L'M\rangle|^{2}}{E_{n'L'M}-E_{nLM}}\,.\label{eq:polarizability}
\end{equation}
 Numerical evaluation of Eq.~(\ref{eq:polarizability}) for the 50F
state shows that the sum over intermediate states is dominated by
a single term, the almost degenerate 50G state. Even in weak fields
the resulting large polarizability leads to sizable shifts in energy
(see Fig.~\ref{fig:stark}). For the ``52P'' state coupling to
the ``52D'' state dominates the sum in Eq.~(\ref{eq:polarizability}).
However, the large energy difference (see Fig.~\ref{fig:stark})
between these states suppresses large values of $\alpha$. The ``52D''
state lies between the two states ``52P'' and ``50F'' to which
it is dipole coupled (Fig.~\ref{fig:stark}). The coupling with the
``50F'' state is dominant leading to polarization on the ``downhill''
side. As the applied field increases, the growth of the dipole moment
becomes non-linear in $F_{{\rm dc}}$ due to the increasing importance
of the higher-order perturbation terms, i.e., the strong mixing with
higher $L$ states. The states can then become strongly polarized
and, as seen for the ``50F'' and ``52D'' states, their polarizations
can approach the limiting value of $\langle z_{1}+z_{2}\rangle=1.5n^{2}$~a.u.

\section{ Conclusions }

The present work demonstrates that very-high-$n$ strontium Rydberg
states with $n$ $\sim$~300 can be excited with remarkable efficiency
by two-photon excitation using a crossed laser-atom beam geometry.
Near simultaneous creation of $\sim$~15 Rydberg atoms within a 500~ns
laser pulse is achieved. In fact, for spectroscopy of isolated strontium
Rydberg atoms it was necessary to reduce either the beam density or
the laser intensity. The photoexcitation spectra recorded in the presence
of a dc field suggest it should be possible to create strongly polarized
Rydberg states by exciting ``$n$D'' states in dc fields approaching
the crossing field $F_{{\rm cross}}$. Interpretation of the data
was facilitated by a full two-active electron approximation in which
the Sr$^{2+}$ core is represented by a semi-empirical core potential
and which allows the calculation of the excitation spectrum both in
the absence and in the presence of a static electric field for Rydberg
states with values $n$ of up to $n\simeq50$. Use of Coulomb scaling
including quantum defects and Ritz corrections allows prediction of
excitation spectra for very high $n$ Rydberg states.

The present measurements and simulations suggest that quasi-one dimensional
polarized strontium states can be formed at $n\simeq300$ that closely
resemble extreme parabolic states. Such quasi-one-dimensional states
provide a valuable starting point for wave function engineering and
can be readily transformed into circular or elliptic wave packets
\cite{dunn09}. Wave packet manipulations open up the opportunity
to explore the creation of quasi-stable two-electron excited states
in which the two electrons are placed in large correlated near-classical
orbits that are stabilized by their mutual interactions~\cite{kali03,eich90,eich92,rich90}.
The capability to create multiple Rydberg atoms during a single laser
pulse will allow the study of strongly-coupled Rydberg-Rydberg systems
at very high $n$. The time evolution of the resulting strongly-correlated
wave packet may open up the possibility, through periodic driving,
of creating a correlated ``molecular'' phase-locked Rydberg wave
packet. 
\begin{acknowledgments}
Research supported by the NSF under Grant No. 0964819, the Robert
A. Welch foundation under Grant No. C-0734, and by the FWF (Austria)
under grant No. P23359-N16. The Vienna Scientific Cluster was used
for the calculations. 
\end{acknowledgments}


\newpage

\setlength{\tabcolsep}{10pt} 
\begin{table}
\centering %
\begin{tabular}{cccccc}
\hline 
Isotope  & Abundance (\%)  & I  & F  & Shift (MHz)  & Relative Strength \tabularnewline
\hline 
$^{84}$Sr  & 0.56  & 0  & -  & $-270.8$  & 1 \tabularnewline
$^{86}$Sr  & 9.86  & 0  & -  & $-124.5$  & 1 \tabularnewline
\vspace*{-0.2cm}
  &  &  & 7/2  & $-9.7$  & 4/15 \tabularnewline
\vspace*{-0.2cm}
 $^{87}$Sr  & 7.00  & 9/2  & 9/2  & $-68.9$  & 1/3 \tabularnewline
 &  &  & 11/2  & $-51.9$  & 2/5 \tabularnewline
$^{88}$Sr  & 82.58  & 0  & -  & $0$  & 1 \tabularnewline
\end{tabular}\caption{Properties of naturally occurring strontium. Isotope shifts and hyperfine
splittings for the 5s$^{2}$ $^{1}$S$_{0}$ $\to$ 5s5p $^{1}$P$_{1}$
transitions are expressed relative to the principal $^{88}$Sr isotope.}
\end{table}

\end{document}